\documentclass[amsmath,amssymb,aps,prb,twocolumn,superscriptaddress,showpacs,floatfix,10pt]{revtex4-1}
\usepackage{mathbbol,amsmath,amsfonts,amssymb,bm,color,ulem}
\usepackage{graphicx,psfrag,array,braket,mathtools,physics,multirow}
\newcommand{\comment}[1]{}

\def\be{\begin{equation}}
\def\ee{\end{equation}}
\def\beq{\begin{eqnarray}}
\def\eeq{\end{eqnarray}}
\begin{document}
\title{Classification of superconducting orders in nonhermitian systems}
\author{Sumanta Bandyopadhyay}
\affiliation{Nordita, KTH Royal Institute of Technology and Stockholm University, Roslagstullsbacken 23, SE-106 91 Stockholm, Sweden}
\author{Alexander Balatsky}
\affiliation{Nordita, KTH Royal Institute of Technology and Stockholm University, Roslagstullsbacken 23, SE-106 91 Stockholm, Sweden}
\affiliation{Department of Physics, University of Connecticut, Storrs, Connecticut 06269, USA}
\date{\today}
\begin{abstract}
We have investigated the effects of nonhermiticity on superconducting states. We have shown the definition of the time-ordered product in the Green's function requires the introduction of a bi-orthogonal basis in nonhermitian systems. This modification allows a classification scheme of nonhermitian superconductors with new classes of states not present in hermitian systems.   Furthermore, we have discussed the density of state profiles and signatures in spin susceptibility measurements for different nonhermitian superconducting orders using the simplest case of BCS superconductors.
\end{abstract}
\maketitle
\section{Introduction}
Since the discovery of superconductivity \cite{onnes1911disappearance} and the development of its microscopic theory \cite{BCS},  we have come across some very exotic phases of matter in superconductors (SC). The new variety of superconducting states, the so-called odd frequency pairing states (OFSc), were first proposed by Berezinskii \cite{berezinskiiOdd}, where anomalous pairing amplitude, ${\sf F}_{\alpha\beta,ab}(x,t;x',t')$ is odd under permutation of time coordinates $t.t'$.
\be\label{eq:1}
{\sf F}_{\alpha\beta,ab}(x,t;x',t')=\langle T_t c_{\alpha a}(x,t)c_{\beta b}(x',t')\rangle
\ee
Using the very definition of the time-ordered ($T_t$) product, one can show anomalous pairing amplitude obey the following symmetry relation.
\be\label{eq:spot}SP^*OT^*{\sf F}_{\alpha\beta,ab}(x,t;x',t')=-{\sf F}_{\alpha\beta,ab}(x,t;x',t')\ee

Where $S$ and $O$ exchange spin ($\alpha\leftrightarrow\beta$) and orbital ($a\leftrightarrow b$) quantum numbers respectively, while $P^*$ and $T^*$ do the same for spatial ($x\leftrightarrow x'$) and temporal ($t\leftrightarrow t'$) co-ordinates \cite{linderRev}. One can then classify the pairing states in terms of  parities of $F_{\alpha \beta, ab}(t,t')$.

The growing realization that OFSc is a dynamic phase of matter, \cite{trioladirven1,trioladirven2, linderRev}, led us to ask the question: what are the consequences of nonhermitian dynamics of SC state for OFSc? In the next section, we will introduce nonhermiticity in physical systems as another key-player in realizing the OFSc in the bulk of dynamic/ open quantum systems.  We will further see that nonhermitian realization opens up the possibility of different kinds of hidden order parameters in superconducting systems. Namely, we find new superconducting order parameters that,  while still satisfying the $SP^*OT^*=-1$ symmetry, are odd under particle-hole symmetry and thus vanish at the Fermi-surface. Due to the nonhermiticity, a particle can decay out of (pumped into) the system, which in turn changes the volume of the Fermi sea and thus introduces non-trivial action of the particle-hole conjugation. In this context, it is important to introduce a new symmetry that we denote as $C^*$  (electron-hole transformation)  that turns electrons into holes and vise versa. 

We will see that $C^*$ has non-trivial action in the definition of time-ordered product in nonhermitian systems. Which, in turn, gives rise to two branches of Berezinskii classification, as shown in Table \ref{tab:class}.
\beq &&SP^*OT^*{\sf F}^{+}_{\alpha\beta,ab}(x,t;x',t')=-{\sf F}^{+}_{\alpha\beta,ab}(x,t;x',t')\\
&&SP^*OT^*{\sf F}^{-}_{\alpha\beta,ab}(x,t;x',t')=-{\sf F}^{-}_{\alpha\beta,ab}(x,t;x',t')\eeq
Where,  $\pm$ in the ${\sf F}^\pm$ denotes the parity of the anomalous Green's function under the action of $C^*$.
\be C^*{\sf F}^\pm_{\alpha\beta,a b}(x,t;x',t')=\pm{\sf F}^\pm_{\alpha\beta,a b}(x,t;x',t')\ee
\begin{table}
\label{tab:class}
 \centering
\def\arraystretch{1.5}
 \begin{tabular}{|c|l|r|r|r|r||c|l|r|r|r|r|}
 \hline
  $~$& \multicolumn{1}{c|}{$S$} & \multicolumn{1}{c|}{$P^*$} & \multicolumn{1}{c|}{$O$} & \multicolumn{1}{c|}{$T^*$} & \multicolumn{1}{c||}{${\sf F}^{C^*}_{SP^*OT^*}$}&$~$& \multicolumn{1}{c|}{$S$} & \multicolumn{1}{c|}{$P^*$} & \multicolumn{1}{c|}{$O$} & \multicolumn{1}{c|}{$T^*$} & \multicolumn{1}{c|}{${\sf F}^{C^*}_{SP^*OT^*}$}\\
 \hline
 \parbox[t]{2mm}{\multirow{4}{*}{\rotatebox[origin=c]{90}{$C^*=+1$}}}
 & $-1$ & $+1$& $+1$& $+1$& ${\sf F}^{+}_{-+++}$& \parbox[t]{2mm}{\multirow{4}{*}{\rotatebox[origin=c]{90}{$C^*=+1$}}}
 & $+1$ & $+1$& $+1$& $-1$& ${\sf F}^{-}_{+++-}$\\
 & $+1$ & $-1$& $+1$& $+1$& ${\sf F}^{+}_{+-++}$& &$-1$ & $-1$& $+1$& $-1$& ${\sf F}^{+}_{--+-}$\\
 & $+1$ & $+1$& $-1$& $+1$& ${\sf F}^{+}_{++-+}$& &$-1$ & $+1$& $-1$& $-1$& ${\sf F}^{+}_{-+--}$\\
 & $-1$ & $-1$& $-1$& $+1$& ${\sf F}^{+}_{---+}$& &$-1$ & $-1$& $+1$& $-1$& ${\sf F}^{+}_{+---}$\\
 \hline\hline
  \parbox[t]{2mm}{\multirow{4}{*}{\rotatebox[origin=c]{90}{$C^*=-1$}}}
 & $-1$ & $+1$& $+1$& $+1$& ${\sf F}^{-}_{-+++}$& \parbox[t]{2mm}{\multirow{4}{*}{\rotatebox[origin=c]{90}{$C^*=-1$}}}
 & $+1$ & $+1$& $+1$& $-1$& ${\sf F}^{-}_{+++-}$\\
 & $+1$ & $-1$& $+1$& $+1$& ${\sf F}^{-}_{+-++}$& &$-1$ & $-1$& $+1$& $-1$& ${\sf F}^{-}_{--+-}$\\
 & $+1$ & $+1$& $-1$& $+1$& ${\sf F}^{-}_{++-+}$& &$-1$ & $+1$& $-1$& $-1$& ${\sf F}^{-}_{-+--}$\\
 & $-1$ & $-1$& $-1$& $+1$& ${\sf F}^{-}_{---+}$& &$-1$ & $-1$& $+1$& $-1$& ${\sf F}^{-}_{+---}$\\
 \hline
 \end{tabular}
 \caption{Classification of superconducting orders in the nonhermitian system: We have two branches of anomalous pairing amplitudes (${\sf F}$) with different parities under particle-hole conjugation operator  $C^*$. Each of these branches satisfies the $SP^*OT^*=-1$ symmetry constraint. We explicitly write the parities of ${\sf F}$ to avoid confusion. ${\sf F}^{+}_{-+++}$, for example, denotes a particle-hole even $(C^*=+1)$, spin-singlet $(S=-1)$, even parity ($P^*=+1$), single orbital $(O=+1)$ and even frequency $(T^*=+1)$ state. Nonhermiticity in BCS ($F^{+}_{-+++}$) state can induce odd frequency ($F^{+}_{+++-}$) and/or particle-hole odd ($F^{-}_{-+++}$) superconducting orders.}
 \end{table}

In the model we study in this paper, $C^*$ will result in exchanging $\epsilon$ to $-\epsilon$. $\epsilon$ is the quasiparticle energy. We thus see a naturally emerging class of superconducting states that are $\epsilon$-odd. Odd $\epsilon$ SC were proposed earlier \cite{cohen1964coherent} and studied extensively in the case of  gapless \cite{nakajima1964superconductivity} and gapped \cite{PhysRevLett.67.2379}  superconductivity. Odd $\epsilon$ pairing arouse in those cases from the specific choice of interaction potential, not from the nonhermitian dynamics. Nonhermiticity induces particle number nonconservation and converts electrons into holes, thus it produces correlations between hole and electron excitations. This conversion implies that $C^* \neq 1$. In the case of the hermitian system, we do not have any change in the Fermi sea volume induced by decay(pump), and thus we would expect the prevalence of  $C^*=1$ states. 

To prove the classification in Table \eqref{tab:class}, we have introduced the idea of time-ordered pairing amplitude in the nonhermitian system. To check our definition of time-ordered pairing amplitude, we have further proved Berezinskii classification as a consequence of topological classification of particle-hole symmetry in the BdG Hamiltonian (PHS-BdG) for hermitian systems. We will show, while PHS-BdG is broken, $SP^*OT^*=-1$  still can be recovered by extending the PHS-BdG formalism to bi-orthogonal particle-hole basis in the nonhermitian system. 

This paper is organized as follows. In sec-II, we introduce nonhermiticity in the FSF junction and produce spin-triplet bulk OFSc. In sec-III, we review some of the basic properties of time-ordered  Green's function in hermitian systems and extend those notions to nonhermitian systems. While maintaining the $SP^*OT^*=-1$ symmetry, the definition of time-ordered anomalous Green's function in the bi-orthogonal basis opens up various new superconducting channels as shown in Table \ref{tab:class}. In sec-IV, we proved the $SP^*OT^*=-1$ symmetry as a consequence of the PHS in BdG construction of the Hamiltonian and thus substantiated our definition of time-ordered product in nonhermitian system. We have further constructed some nonhermitian superconducting channels from a leaky BCS superconductor and constructed their density of state profiles in Sec-V. In Sec-VI, we have studied Pauli spin-susceptibility, which gives rise to distinct signatures in the nonhermitian BCS superconductors. In conclusion, we have summarized the implications of our classification schemes, which widens the scope of realizing more exotic superconductors in the dynamic systems. 
\section{Odd frequency superconductivity in nonhermitian system}
\label{secII}
Nonhermitian dynamics appears as a result of coupling to the environment. Instead of studying the Hamiltonian $H_0$ of an isolated system, we use the retarded Green's function,
\begin{equation}
    {\sf G}(\omega)=(\omega-H_0-\Sigma(\omega))^{-1}=(\omega-H_{eff}(\omega))^{-1},
\end{equation}
where $\Sigma(\omega)$ is the self-energy term. The environment, which is a continuum spectrum of energy, can have a contribution to the self-energy even at $\omega=0$. The poles of the $G(\omega)$ will serve as nonhermitian effective Hamiltonian with complex eigenvalues \cite{avila2019non}.
\begin{equation}
H_{eff}(\omega)=H_0+\Sigma(\omega), ~\text{s.t, } H_{eff}=H_0-i \Gamma 
\end{equation}

\begin{figure}
    \centering
    \includegraphics[scale=.16]{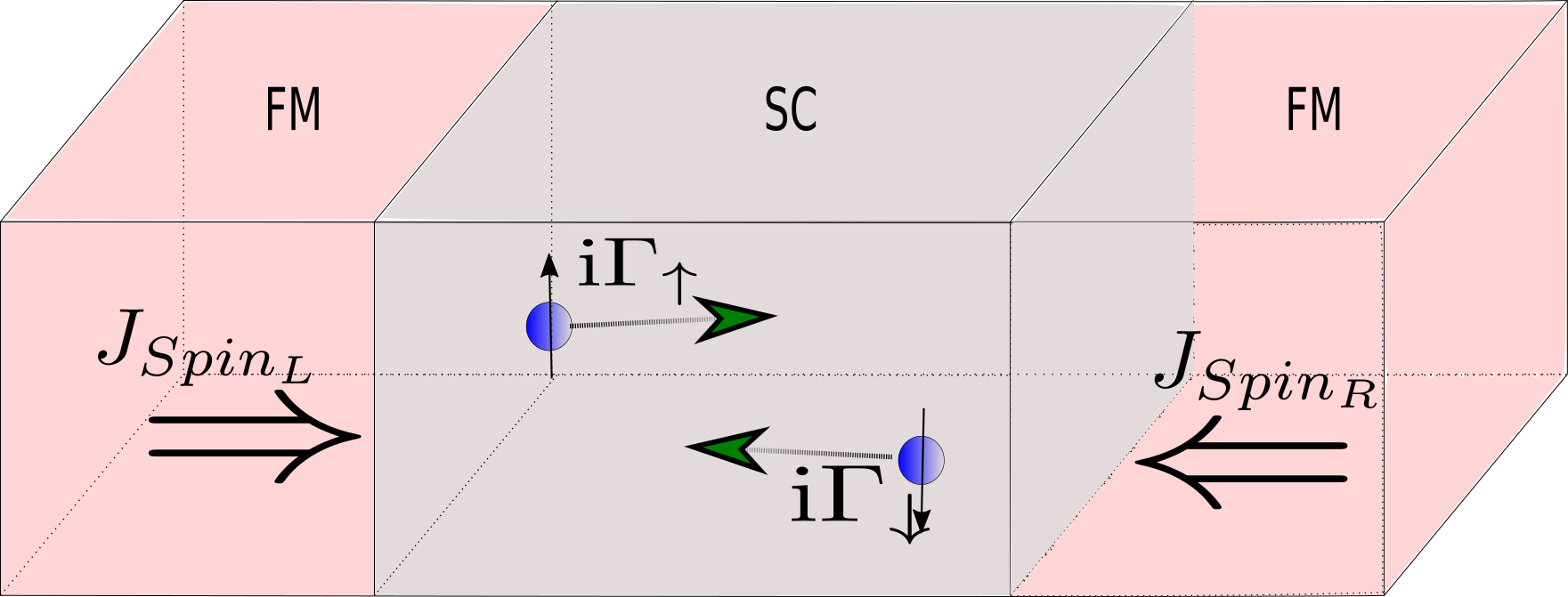}
    \caption{Nonhermitian BdG Hamiltonian can be realized in the FSF junction. The spin-dependent current from the ferromagnets can result in different lifetimes for spin-up ($\Gamma_{\uparrow}$) and spin-down  ($\Gamma_{\downarrow}$) electrons in the superconductor.}
    \label{fig:SFS}
\end{figure}

To realize the OFSc, we will follow a similar mechanism in a ferromagnetic-superconductor-ferromagnetic (FSF) junction (Fig. \ref{fig:SFS}). 
Let us assume, the opposite sides of a bulk superconductor are connected to two FM reservoirs with opposite spin alignments. Without loss of generality, let us assume the left FM has a spin-up orientation.  If we introduce a small current flow from right to left, spin-up electrons will enter the bulk of the superconductor, while spin-down electrons will leave the bulk. Thus, we get finite lifetime, $i\Gamma_\sigma$ for $\sigma=\uparrow,\downarrow$ spins. For this particular scenario, we get, $\Gamma_\uparrow=-\Gamma_\downarrow=\Gamma$.  Let us assume the bulk superconductor to be BCS type superconductor i.e, a spin-singlet, s-wave single orbital superconductor. In the isolated system, we expect the anomalous pairing amplitude to satisfy the following relation,
\begin{equation*}
    \langle T_t c_{\downarrow}(x,t)c_{\uparrow}(x't')\rangle_0=-\langle T_t c_{\uparrow}(x,t)c_{\downarrow}(x't')\rangle_0
\end{equation*}
$\langle\rangle_0$ stands for the isolated system. Now, in the original nonhermitian system with finite lifetime for the spins, we expect,
\begin{equation*}
    \langle T_t c_{\sigma}(x,t)c_{\sigma'}(x't')\rangle\sim e^{-\Gamma_\sigma t-\Gamma_\sigma't'}\langle T_t c_{\downarrow}(x,t)c_{\uparrow}(x't')\rangle_0
\end{equation*}
For our particular choice of $\Gamma_\sigma$s, we get, 
\begin{eqnarray}\nonumber
{\sf F_s}=\langle T_t c_{\downarrow}(x,t)c_{\uparrow}(x't')\rangle-\langle T_t c_{\uparrow}(x,t)c_{\downarrow}(x't')\rangle\\=\cosh(\Gamma(t-t'))\langle T_t c_{\downarrow}(x,t)c_{\uparrow}(x't')\rangle_0\\\nonumber
{\sf F_t}=\langle T_t c_{\downarrow}(x,t)c_{\uparrow}(x't')\rangle+\langle T_t c_{\uparrow}(x,t)c_{\downarrow}(x't')\rangle\\=\sinh(\Gamma(t-t'))\langle T_t c_{\downarrow}(x,t)c_{\uparrow}(x't')\rangle_0
\end{eqnarray}
For the finite lifetime of electron spins, we thus expect nonhermitian dynamics inducing  spin-triplet odd frequency (Berezinskii pairing, $S = 1$, $P^* = 1$, $O = 1$ (single band), $ T^* = -1$) in addition of the spin-singlet BCS pairing present in the isolated system. This example vividly demonstrates the role of the nonhermitian term play in generating Berezinskii pairing. The OFSc is a dynamic order. Nonhermitian terms imply the changes in the state of the system with time and induce its own dynamics. As a result, OFSc is induced.  We thus surmise that  OFSc be induced on general grounds in any nonhermitian superconducting system.
\section{Nonhermitian quantum mechanics}
\label{secIII}
We have, so far, convinced ourselves that nonhermitian dynamics can introduce OFSc in the bulk BCS superconductor. Introduction of nonhermiticity, however, poses another vital question: Do nonhermitian systems obey Berezinskii classification? To answer this, we must familiarize ourselves with the subtleties and their implications in the nonhermitian world. We find it beneficial to start our discussion with a brief review of hermitian systems and extend the correspondence to the nonhermitian world.
\subsection{Hermitian system in equlibrium}
A hermitian system with Hamiltonian $H$ can be described by Schrodinger equation,
\begin{equation*}
(i\partial_t-H)\ket{\psi_\alpha(x,t)}=0;~~~~\ket{\psi_{\alpha}(x,t)}=\psi_\alpha^\dagger(x,t)\ket{0}
\end{equation*}
Where, $\ket{0}$ is the ground state of the system under discussion. $\psi_\alpha^\dagger(x,t)$ creates a particle at position $x$, time $t$ and quantum number $\alpha$, while  $\psi_\alpha(x,t)$ destroys it. Moreover, $\psi_\alpha(x,t)$ and $\psi_\alpha^\dagger(x,t)$ are related by hermitian conjugation. $\psi_{\alpha}(x,t)$ forms a complete and orthogonal basis for the Hamiltonian $H$, as
\be\sum_{\alpha}\psi(x.t)\psi_{\alpha}^{\dagger}(x',t)=\delta^{(d)}(x-x')\ee
Where $d$ is the spatial dimension. One can define the corresponding Dyson's equation,
\begin{equation}\label{eq:dy}
(i\partial_t-H){\sf G}(x,t;x',t')=\delta^{(d)}(x-x')\delta(t-t')
\end{equation}
Where the Green's function matrix ${\sf G}(x,t;x',t')$ is defined in terms of time-ordered product,
\begin{equation*}
    {\sf G}(x,t;x',t')=\theta_{tt'}{\sf G}^>(x,t;x',t')-\theta_{t't}({\sf G}^<(x',t';x,t))^T
\end{equation*}
$^T$ in the last term stands for transpose of the matrix ${\sf G}^<(x,t;x',t')$. $\theta_{tt'}$ is the Heavyside-theta function,
\begin{equation*}
    \theta_{tt'}=1\text{ if }t>t', 0\text{ if }t<t'
\end{equation*}
Matrix elements of $G^{^\lessgtr}(x,t;x',t')$ are given in terms of ensemble average $\langle\rangle$,
\beq\nonumber
{\sf G}^<_{\alpha\beta}(x,t;&x',t')=&\langle{\psi_{\alpha}(x,t)}{\psi^\dagger_{\beta}(x',t')}\rangle\\\nonumber
    =\tr&[\psi_{\alpha}(x,t)&\psi_{\beta}^\dagger(x',t')\rho(0)]\\\nonumber
{\sf G}^>_{\alpha\beta}(x,t;&x',t')=&\langle{\psi_{\alpha}^\dagger(x,t)}{\psi_{\beta}(x',t')}\rangle\\\nonumber
    =\tr&[\psi_{\alpha}^\dagger(x,t)&\psi_{\beta}(x',t')\rho(0)]\\
    \implies{\sf G}_{\alpha\beta}(x,t;&x',t')=&\langle T_t\psi_{\alpha}(x,t)\psi_{\beta}^\dagger(x',t')\rangle\label{eq:gf}\\\nonumber=\theta_{tt'}\langle\psi_{\alpha}(x,t)&\psi_{\beta}^\dagger(x',t')&\rangle-\theta_{t't}\langle\psi_{\beta}^\dagger(x',t')\psi_{\alpha}(x,t)\rangle  
\eeq
Where $T_t$ stands for time ordered product,  $\alpha,\beta=1,...,N$ are good quantum numbers and $\rho(0)$ is the density matrix at $t=0$, given by,
\be \rho(t)=\sum_{\alpha}\int d^dx \ket{\psi_\alpha(x,t)}\bra{\psi_{\alpha}(x,t)}\ee
From the Schrodinger equation, we get,
\be
\frac{d}{dt}\rho(t)=-i[H,\rho(t)]\implies\rho(t)=e^{-i Ht}\rho(0)e^{i Ht}
\ee
The above equation implies the operators in the hermitian system have unitary time evolution. As a direct consequence to this unitary time evolution, we get norm conservation property, i.e,
\be\tr[\rho(t)]=\tr[\rho(0)],\ee
which suggests the total particle number is conserved. Hence, we have two equivalent definitions for Green's functions, namely, Eq. \eqref{eq:dy} and Eq. \eqref{eq:gf}.For an isolated system with Hamiltonian $H$, $G_{\alpha\beta}(x,t;x',t')$ is analytic everywhere except $t=t'$. For $t=t'$, we bypass the light-cone singularity by introducing analytic continuity in terms of time-ordered product. The analyticity of the Green's function in the hermitian system is often used to construct observables, such as the density of states. The ab-initio definition of the density of states is given by \cite{feinberg1997non},
\be\rho_E(\omega,\omega_1)=\langle\frac{1}{N}\sum_{i}\delta(\omega-\Re{E_i})\delta(\omega_1-\Im{E_i})\rangle \label{eqnhdos} \ee
$E_i$s are eigenvalues of the Hamiltonian $H$ in $N\times N$ matrix representation. In the hermitian case, eigenvalue being always real,  we often drop the imaginary part of the eigenvalue from the definition of the density of states,
\be\rho_E^h(\omega)=\langle\frac{1}{N}\sum_i\delta(\omega-E_i)\rangle\ee
$^h$ stands for the definition of density of states in the hermitian system. Using the analytic properties of the Green's function, we get,
\be\rho_E^h(\omega)=\frac{1}{\pi}\lim_{\eta\rightarrow0+}\langle\Im{{\sf G}(\omega-i\eta)}\rangle\ee

$\sf G(\omega)$ can be derived by integrating ${\sf G}(k,\omega)$ over all values of momentum $k$. ${\sf G}(k,\omega)$ is the Fourier transform of ${\sf G}(x,t;x',x',t')$ in the bulk of the system, s.t, 
\be(\omega-H(k)){\sf G}(k,\omega)=1\ee

The latter definition of $\rho^h_E(\omega)$ in terms of Green's function is more useful in terms of computational ease. Moreover, this notion of the density of states can be generalized to interacting systems, where a Hamiltonian description is not readily available. In the remaining part of this section, we will primarily extend these Hamiltonian and Green's function formalism in the nonhermitian realm. 
\subsection{Nonhermitian system}
Analogous to a hermitian system in equilibrium, we will construct a Hamiltonian/ Green's function description for nonhermitian systems. The following construction is performed in real time but can be straightforwardly extended to a Matsubara time formalism. 
In sec-\ref{secII}, we have established that an open/ dynamic system can have an effective nonhermitian description. This effective description seems to imitate an isolated system with a Hamiltonian, accompanied by the following subtleties: Hamiltonian, which is no longer hermitian, can have complex eigenvalues and eigenvectors might not give a complete basis description. We will start with a simple single-particle Hamiltonian $H$ to lead the discussion on the impact of these subtleties.
\begin{equation}
    H= \sum_\alpha \epsilon_\alpha c^\dagger_\alpha c_\alpha
\end{equation}
A fermion with quantum number $\alpha$ is destroyed by annihilation operator $c_\alpha$ and created by $c^\dagger_\alpha$. The matrix element related to this process is given by $\epsilon_\alpha$. Now, let us assume $\epsilon_\alpha$ is not hermitian and $c^\dagger_\alpha$ is hermitian conjugate of $c_\alpha$. If we take hermitian conjugate of the above Hamiltonian, we get,
\begin{equation}
    H^\dagger=\sum_\alpha \epsilon^\dagger_\alpha c^\dagger_\alpha c_\alpha
\end{equation}
$H$ and $H^\dagger$ describe a similar process with different matrix elements $\epsilon_\alpha$ and $\epsilon^\dagger_\alpha$, thus introducing contradiction to our previous assumptions. To resolve this issue, we assume that the creation and annihilation operations are not hermitian conjugates of each other. A proper representation of the nonhermitian Hamiltonian would be,
\begin{equation}\label{eq:nh}
    H=\sum_\alpha \epsilon_\alpha \Tilde{c}^\dagger_\alpha
    c_\alpha;~~~~H^\dagger=\sum_\alpha \epsilon_\alpha c^\dagger_\alpha \Tilde{c}_\alpha;
\end{equation}
Where $\dagger$ stands for the usual Hermitian conjugation but $c_{\alpha}$ and $c^\dagger_\alpha$  no longer annihilate and create the same particle. We can give the following interpretation for these operators, summed up in Table \ref{table11}.  
\begin{table}
\begin{tabular}{|l|l|}
\hline
$c_{\alpha}$ & electron destruction operator \\ \hline
$\Tilde{c}_{\alpha}$ & hole creation operator \\ \hline
$c^\dagger_{\alpha}$ & hole destruction operator \\ \hline
$\Tilde{c}^\dagger_\alpha$ & electron creation operator \\ \hline
\end{tabular}
\caption{A nonhermitian system can be given an effective hermitian description in the bi-orthogonal basis. Hole creation and electron destruction are given by two different operations, $c_{a}$, $\Tilde{c}_{a}$ respectively.}\label{table11}
\end{table}
Hence, we get a matrix description of the Hamiltonian $H$ in terms of $c_\alpha$, $\tilde{c}^\dagger_\alpha$ with the following non-trivial anti-commutation relations,
\be\{c_{\alpha},\tilde{c}^\dagger_{\beta}\}=\{\tilde{c}_{\alpha},{c}^\dagger_{\beta}\}=\delta_{\alpha\beta};\ee 
Rest of the anti-commutation relations are equal to zero. 
Using the Eq. \eqref{eq:nh}, we would like to define the following Green's functions in terms of ensemble average,
\begin{eqnarray}
    {\sf G}^<_{\alpha\beta}(t,t')=\langle c_{\alpha}(t)\tilde{c}^\dagger_{\beta}(t')\rangle=\tr[c_{\alpha}(t)\tilde{c}^\dagger_{\beta}(t')\rho(0)]\\
    {\sf G}^>_{\alpha\beta}(t,t')=\langle c_{\alpha}^\dagger(t)\tilde{c}_{\beta}(t')\rangle=\tr[c_{\alpha}^\dagger(t)\tilde{c}_{\beta}(t')\rho(0)]
\end{eqnarray}
We have dropped $x,x'$ from the Green's functions as the spatial dependence does not get affected by non-hermiticity.  Density matrix $\rho(t)$ evolves according to the following equation,
\be\frac{d}{dt}\rho(t)=-\frac{i}{2}[H+H^\dagger,\rho(t)]+\frac{i}{2}\{H-H^\dagger,\rho(t)\}\ee

Where $[,]$ and $\{,\}$ stand for commutation and anti-commutation relation respectively. The above equation no longer allows a unitary solution. We rather have,
\be\rho(t)=e^{-iHt}\rho(0)e^{iH^\dagger t}\implies\frac{d}{dt} \tr[\rho(t)]\ne 0\ee

The above equation highlights the fact that the particle number in the non-hermitian system can change over time. Hence, the density matrix may no longer be normalized to identity in two different times. This observation prompts us to modify ${\sf G}^{^\lessgtr}(t,t')$ in the following way,
\begin{eqnarray}
    {\sf G}^<_{\alpha\beta}(t,t')=\langle c_{\alpha}(t)\tilde{c}^\dagger_{\beta}(t')\rangle=\frac{\tr[c_{\alpha}(t)\tilde{c}^\dagger_{\beta}(t')\rho(0)]}{\tr[\rho(t)]}\\
    {\sf G}^>_{\alpha\beta}(t,t')=\langle c_{\alpha}^\dagger(t)\tilde{c}_{\beta}(t')\rangle=\frac{\tr[c_{\alpha}^\dagger(t)\tilde{c}_{\beta}(t')\rho(0)]}{\tr[\rho(t)]},
\end{eqnarray}
where we have normalized ${\sf G}^{^\lessgtr}(t,t')$ at time $t$. We can define the time-ordered Green's function, 
\beq \nonumber {\sf G}_{\alpha\beta}(t,t')=\theta_{tt'}{\sf G}_{\alpha\beta}^>(t't)-\theta_{t't}{\sf G}_{\beta\alpha}^<(t',t)\\=\frac{\tr[c_{\alpha}(t)\tilde{c}^\dagger_{\beta}(t')\rho(0)]}{\tr[\rho(t)]}-\frac{\tr[c_{\beta}(t')^\dagger\tilde{c}_{\alpha}(t)\rho(0)]}{\tr[\rho(t')]}\eeq
Time-ordered Green's function plays a central role in the hermitian system. There ${\sf G}(t,t')$ is the solution of the response function for the linearized equation of motion. Does ${\sf G}(t,t')$ have a similar role in the nonhermitian physics? Is it related to physical observables, such as the density of states? Do eigenstates of the Hamiltonian spans a complete basis set? In other words, does ${\sf G}(t,t')$ constitute the solution for the following equation?
\be (i\partial_t-H)G(t,t')=\delta(t-t') \label{eq0} \ee
To answer these subtleties, we will start with an auxiliary problem,
\be (i\partial_t-H^\dagger)\Tilde{G}(t,t')=\delta(t-t') \label{eq1} \ee
Instead of studying Eq. \eqref{eq0}-\eqref{eq1} separately, we should study them together in the bi-orthogonal basis \cite{curtright2007biorthogonal,brody2013biorthogonal} $\psi=(\Tilde{c},c)$. The first $N$ rows of $\psi$ are electron annihlation operators $c_{\alpha}$s and last $N$ rows are hole creation operators $\Tilde{c}_{\alpha}$s  for $\alpha=1,...,N$. We have thus doubled up the basis of the problem but the resulting system has a hermitian Hamiltonian $\mathcal{H}$ in the basis of $\psi$,
\be\mathcal{H}=\begin{pmatrix}0&H\\H^\dagger&0\end{pmatrix}\ee

We map this hermitian Hamiltonian to the original nonhermitian problem by introducing an operator $C^*$, which connects the particle and hole sectors,
\be C^* c_{\alpha}=\tilde{c}_{\alpha}; ~~~C^* c^\dagger_{\alpha}=\tilde{c}^\dagger_{\alpha};~~~C^{*^2}=1\ee
In a hermitian system $C^*=1$,  but in general it can take both $\pm 1$ values. The response function for $\psi$ can written as,
\be\lim_{\eta\rightarrow0+}\begin{pmatrix}\sqrt{\eta}& (i\partial_t-H)\\(i\partial_t-H)^\dagger&-\sqrt{\eta}\end{pmatrix}\mathcal{G}(t,t')=\delta(t-t')\label{eq2} \ee
Where, 
\begin{equation*}
\mathcal{G}(t,t')=\langle T_t\psi(t)\psi^\dagger(t')\rangle=\begin{pmatrix}\langle T_t\Tilde{c}(t)\Tilde{c}^\dagger(t')\rangle&\langle T_t\Tilde{c}(t){c}^\dagger(t')\rangle\\\langle T_t {c}(t)\Tilde{c}^\dagger(t')\rangle&\langle T_t{c}(t){c}^\dagger(t')\rangle\end{pmatrix}
\end{equation*}
But we have already defined, 
${\sf G}_{\alpha\beta}(t,t')=\langle T_t\Tilde{c}_{\alpha}(t){c}_{\beta}^\dagger(t')\rangle$
Let us define a matrix ${\sf G}(t,t')$ with matrix elements ${\sf G}_{\alpha,\beta}(t,t')$ and another matrix, $\Tilde{\sf G} (t,t')$ with matrix elements, 
\be\Tilde{\sf G}_{\alpha\beta}(t,t')=\langle T_t{c}_{\alpha}(t)\Tilde{c}_{\beta}^\dagger(t')\rangle\ee

Hence, we can rewrite $\mathcal{G}(t,t')$ as,
\begin{equation}
\mathcal{G}(t,t')=\begin{pmatrix}\langle T_t\Tilde{c}(t)\Tilde{c}^\dagger(t')\rangle&{\sf G}(t,t')\\\Tilde{\sf G}(t,t')&\langle T_t{c}(t){c}^\dagger(t')\rangle\end{pmatrix}
\end{equation}
Fourier transforming Eq. \eqref{eq2}, we get,
\be\begin{pmatrix}(\omega-H){\sf G}(\omega)&0\\0&(\omega-H)^\dagger\Tilde{\sf G}(\omega) \end{pmatrix}=\begin{pmatrix}1&0\\0&1\end{pmatrix}\ee
We component-wise Fourier transform the above equation, to show that time-ordered Green's function indeed satisfies the Dyson's equation in a nonhermitian system.
\beq
(i\partial_t-H){\sf G}(t,t')=\delta(t-t');\\(i\partial_t-H)^\dagger\Tilde{\sf G}(t,t')=\delta(t-t');
\eeq
Along with $\langle T_t\Tilde{c}(t)\Tilde{c}^\dagger(t')\rangle$=$\langle T_t{c}(t){c}^\dagger(t')\rangle$=0. Although the above equation looks innocuously similar to its Hamiltonian counterpart, they have much more subtleties \cite{heiss2015green}. For starters, these equations no longer have analytic structures. Eigenvectors of $H$ may not span the complete basis. To ensure the complete basis for Dyson's equation, one must go to the bi-orthogonal basis, where Dyson's equation for ${\sf G}(t,t')$ will always be accompanied by Dyson's equation for $\Tilde{\sf G}(t,t')$. Moreover, eigenvalues of $H$ are no longer real, which implies ${\sf G}(\omega)=\frac{1}{\omega-H}$ can have poles all over the complex plane. As an immediate consequence,  the density of states can no longer be defined in terms of $\Im {\sf G(\omega)}$. Using the ab-initio definition of density of states in Eq. \eqref{eqnhdos}, along with a mathematical identity,
\be\partial_{\omega^*}\frac{1}{\omega}=\pi\delta(\frac{\omega+\omega^*}{2})\delta(\frac{\omega-\omega^*}{2i})\ee
We can define a two-dimensional density of states\cite{feinberg1997non,Chalker}, 
\begin{equation}
    \rho_E(\omega,\omega_1)=\langle\partial_{\tilde{\omega}^*}{\sf G}(\tilde{\omega})\rangle;
\end{equation}
Where $\tilde{\omega}=\omega+i\omega_1$, and $\omega^*$ is complex conjugate of $\omega$. In section-VI, we will discuss the density of states profile for some particular nonhermitian models.
\subsection{Classification of anomalous Green's function}
We have, so far, given the definition of time-ordered Green's function in the nonhermitian system, 
\beq\nonumber
{\sf G}_{\alpha\beta}(x,t;x',t')=\langle T_t{c}_{\alpha}(x,t)\Tilde{c}_{\beta}^\dagger(x',t')\rangle\\\nonumber=\theta_{tt'}\frac{\tr[{c}_{\alpha}(x,t)\tilde{c}_{\beta}^\dagger(x',t')\rho(0)]}{\tr[\rho(t)]}\\-\theta_{t't}\frac{\tr[\tilde{c}_{\beta}^\dagger(x',t')c_{\alpha}(x,t)\rho(0)]}{\tr[\rho(t')]}  
\eeq
In order to establish this time-ordered Green's function as the solution of the nonhermitian Dyson's equation, we have invoked the bi-orthogonal basis. In this basis, we got additional Green's functions, namely, $\langle T_t\tilde{c}_{\alpha}(x,t)\Tilde{c}_{\beta}^\dagger(x',t')\rangle$, $\langle T_t{c}_{\alpha}(x,t){c}_{\beta}^\dagger(x',t')\rangle$ and $\langle T_t\Tilde{c}_{\alpha}(x,t){c}_{\beta}^\dagger(x',t')\rangle$. While the last Green's function satisfy the hermitian conjugate of the Dyson's equation, the other two Green's functions become zero. 

Similarly,  in nonhermitian superconducting systems, we get multiple anomalous paring channels, namely, $\langle T_t{c}_{\alpha}(x,t)\Tilde{c}_{\beta}(x',t')\rangle$, $\langle T_t\tilde{c}_{\alpha}(x,t)\Tilde{c}_{\beta}(x',t')\rangle$, $\langle T_t\tilde{c}_{\alpha}(x,t){c}_{\beta}(x',t')\rangle$ and $\langle T_t{c}_{\alpha}(x,t){c}_{\beta}(x',t')\rangle$. Following the definition of time ordered product, each of these anomalous pairing amplitudes will satisfy the $SP^*OT^*=-1$ rule. However, which of these anomalous pairing amplitude have physical meaning will depend on the details of the interaction potential. For example, $\Tilde{c}^\dagger\Tilde{c}^\dagger cc$ type interaction will give rise to non-zero $\langle T_t{c}_{\alpha}(x,t){c}_{\beta}(x',t')\rangle$ and $\langle T_t\tilde{c}_{\alpha}(x,t)\Tilde{c}_{\beta}(x',t')\rangle$ terms, while $\Tilde{c}^\dagger{c}^\dagger \tilde{c}c$ type interaction will give rise to $\langle T_t{c}_{\alpha}(x,t)\Tilde{c}_{\beta}(x',t')\rangle$ $$\langle T_t\tilde{c}_{\alpha}(x,t){c}_{\beta}(x',t')\rangle$$ terms. In order to incorporate these channels in the classification scheme, we have to introduce an orthogonal $C^*$ axis in the classification schemes, as shown in Table \ref{tab:class}. We will refer to this classification scheme as, $C^*\otimes(SP^*OT^*=-1)$ symmetry.
In the next section, we will elaborate on $C^*$ symmetry as a topological classifier in BdG type Hamiltonian and establish the time-ordered Green's function and $SP^*OT^*=-1$ symmetry as a consequence of BdG construction. 

\section{BdG particle-hole symmetry and Berezinskii classification}
\label{secIV}
In the last section, we have given a consistent definition of the time-ordered product for nonhermitian Green's functions. Our definition allows $SP^*OT^*=-1$ symmetry in the anomalous pairing amplitude even in the nonhermitian system. $SP^*OT^*=-1$ symmetry, however, can also be constructed without explicitly specifying the structure of time-ordered product. In this section, we will give an alternate derivation of the $SP^*OT^*$ rule using the Hamiltonian/bulk Green's function framework. We  start our discussion with the BdG description of the following superconducting Hamiltonian,
\begin{equation}
H= \sum_{k,a,b}c^\dagger_{k,a}H_{a,b}(k)c_{k,b} +\sum_{k,a,b}\Delta_{a,b}(k)c^\dagger_{k,a}c^\dagger_{-k,b}+h.c.   
\end{equation}
Where $a,b=1,...,N$ stand for good quantum numbers, such as spin, orbital quantum number, chirality, etc.
Hence, the corresponding BdG Hamiltonian can be written as, $2N\times2N$ matrix

\begin{equation}
    H_{BdG}(k)=\begin{pmatrix}
    H(k) & \Delta(k)\\-\Delta^*(-k)& -H^T(-k)
    \end{pmatrix}
\end{equation}
In the basis of $\Psi(k)=(c_{k},c^\dagger_{-k})^T$, a column matrix of length $2N$. First half of the $\psi$ is comprised of electron annihilation operators $c_{ka}$ and latter half is made of electron creation operator $c^\dagger_{-ka}$ for $a=1,..,N$. Notice that $H_{BdG}$ will always satisfy the following equation,
\begin{equation}\label{eq:BdGPHS}
    \mathcal{C}H^*_{BdG}(k)\mathcal{C}^{-1}=-H_{BdG}(-k);\quad \mathcal{C}=\begin{pmatrix}0 &I\\I&0
    \end{pmatrix}
\end{equation}
Where $I$ is $N\times N$ identity matrix. The above equation is the artifact of built-in particle-hole symmetry in the BdG Hamiltonian \footnote{This particle-hole symmetry is not related to the physical particle-hole symmetry of the superconductor. It is rather a signature or topological classifier of the hermiticity of the Hamiltonian. In future work, we will discuss the importance of `physical' particle-hole symmetry in OFSc}. Now,
\begin{equation}
(i\partial_t-H_{BdG}(k))G_{BdG}(k,t,t')=\delta(t-t')
\end{equation}
We can use the above definition of the Green's function along with Eq. \eqref{eq:BdGPHS} to infer,
\begin{equation}\label{eq:bdggreen}
    \mathcal{C} G_{BdG}^T(k,t,t')\mathcal{C}^{-1}=-G_{BdG}(-k,t',t)
\end{equation}
Where, $G_{BdG}(k,t,t')$ is hermitian matrix and $G_{BdG}^*(k,t,t')=G_{BdG}^T(k,t,t')$. At this point, it is instructive to write down $G_{BdG}(k,t,t')$ explicitly, 

\beq\nonumber\label{eq:greenMat}
    && G_{BdG}(k,t,t')=\begin{pmatrix} \langle{T_t c_{k}(t)c^\dagger_{k}(t')\rangle} & \langle{ T_tc_{k}(t)c_{-k}(t')}\rangle\\
\langle{T_tc^\dagger_{-k}(t)c^\dagger_{k}(t')}\rangle& \langle{T_tc^\dagger_{-k}(t)c_{-k}(t')}\rangle
    \end{pmatrix}
\eeq
Where, each $\langle\rangle$ stands for an $N\times N$ matrix. s.t,
\be\langle{T_t c_{k}(t)c^\dagger_{k}(t')\rangle}_{ab}=\langle{T_t c_{ka}(t)c^\dagger_{kb}(t')\rangle};\ee
\be\langle{T_t c_{k}(t)c_{-k}(t')\rangle}_{ab}=\langle{T_t c_{ka}(t)c_{-kb}(t')\rangle};\ee
Where $a,b=1,...,n$. 
From the diagonal-blocks, we get, 
\be\langle T_tc_{k,a}(t)c^\dagger_{k,b}(t')\rangle=-\langle T_tc^\dagger_{k,b}(t')c_{k,a}(t)\rangle\ee
This is the definition of time-ordered Green's function. We have thus established that PHS in BdG construction of a hermitian Hamiltonian is equivalent to the time-ordered Green's function. Similarly, we can look into the anomalous pairing channels in the off-diagonal blocks, 
\begin{equation}
{\sf F}(k,t,t')=-{\sf F}^T(-k,t',t)
\end{equation}
Where ${\sf F}(k,t,t')$ is the anomalous Green's function, $\langle{ T_tc_{k}(t)c_{-k}(t')}$. The above equation is just the matrix form of the Berezinskii constraint, Eq. \eqref{eq:spot} in the bulk of the system,
\begin{equation}
{\sf F}_{ab}(k,t,t')=-{\sf F}_{ba}(-k,t',t)
\end{equation}

We now are ready to derive a similar constraint for the nonhermitian system. As predicted in the last section, the constraint is going to be equivalent to the same Berezinskii constraint as for hermitian systems. To see this result, we introduce a bi-orthogonal basis description. We will start with the following nonhermitian Hamiltonian:

\begin{eqnarray}\nonumber
&&H_{nh}=\sum_{k,a,b}\tilde{c}^\dagger_{k,a}(H_{a,b}(k)-i\Gamma_{a,b}(k))c_{k,b}\\&& +\sum_{k,a,b}\Delta_{a,b}(k)\tilde{c}^\dagger_{k,a}\tilde{c}^\dagger_{-k,b}+\Delta^{*}_{a,b}(k) c_{k,a} c_{-k,b} 
\end{eqnarray}
Where, $H(k), \Gamma(k)$ are hermitian matrices.  Now, we can define a BdG biorthogonal basis. 

\begin{equation}
\Psi(k)= (\tilde{c}_{k},{c}^\dagger_{-k},c_{k},\tilde{c}^\dagger_{-k})^T
\end{equation}
$\psi(k)$ is a column matrix with length $4N$. First and last quarters of $\psi$ are comprised of hole creation and particle creation operators $\tilde{c}_{ka}$  and $\tilde{c}^\dagger_{-ka}$, respectively for some ordered set $a=1,..,N$. In between them, we have $N$ hole creation and $N$ particle creation operators, $c^{\dagger}_{-ka}$ and $c_{ka}$.  
In this basis, we can have the following BdG Hamiltonian,
\be H_{BdG}(k)=\begin{pmatrix}0& H_{BdG}^{nh}(k)\\H_{BdG}^{nh^\dagger}(k)&0\end{pmatrix}\ee
$H_{BdG}^{nh}(k)$ is the following nonhermitian matrix.
\begin{equation*}
    H_{BdG}^{nh}(k)=\begin{pmatrix}
    H(k) - i \Gamma(k)& \Delta(k)\\-\Delta^*(-k)& -H^T(-k)+i \Gamma^T(-k)\end{pmatrix}
\end{equation*}
Hence, we will get, 
\begin{equation}\label{Eq:bdgnonherm}
    \mathcal{C} H^{nh^*}_{BdG}(k)\mathcal{C}^{-1}=-H^{nh^\dagger}_{BdG}
\end{equation}

Now, we can again use the Dyson equations,
\beq\label{eq:nhGr}(i\partial_t-H^{nh}_{BdG}(k))G_{BdG}(k,t,t')=\delta(t-t')\\(i\partial_t-H^{nh^\dagger}_{BdG}(k))\tilde{G}_{BdG}(k,t,t')=\delta(t-t')\eeq
along with Eq. \eqref{Eq:bdgnonherm} to deduce,
\beq
\mathcal{C}G_{BdG}^*(k,t,t')\mathcal{C}^{-1}=-\tilde{G}_{BdG}(-k,t',t)\\
\mathcal{C}\tilde{G}_{BdG}^*(k,t,t')\mathcal{C}^{-1}=-{G}_{BdG}(-k,t',t)
\eeq
Where,
\be\nonumber
{G}_{BdG}(k,t,t')=\begin{pmatrix} \langle{T_t {c}_{k}(t)\Tilde{c}^\dagger_{k}(t')}\rangle & \langle{ T_tc_{k}(t)c_{-k}(t')}\rangle\\
\langle{T_t\Tilde{c}^\dagger_{-k}(t)\Tilde{c}^\dagger_{k}(t')}\rangle& \langle{T_t\Tilde{c}^\dagger_{-k}(t)c_{-k}(t')}\rangle
    \end{pmatrix}
\ee
${\Tilde{G}}_{BdG}(k,t,t')$ can be constructed by replacing $c_{k}$ by $\Tilde{c}_{k}$ and vice-versa.

Let us now introduce a bigger matrix.
\be\mathcal{G}_{BdG}(k,t,t')=\begin{pmatrix}0& \tilde{G}_{BdG}(k,t,t')\\G_{BdG}(k,t,t')&0\end{pmatrix}\ee
With this matrix, we can write down a compact equation,
\begin{equation}
    \begin{pmatrix} 0 &\mathcal{C}\\\mathcal{C}&0
    \end{pmatrix} \mathcal{G}_{BdG}^*(k,t,t')
    \begin{pmatrix} 0 &\mathcal{C}^{-1}\\\mathcal{C}^{-1}&0
    \end{pmatrix}= - \mathcal{G}_{BdG}(-k,t',t)
\end{equation}

Now, using the hermiticity of the $\mathcal{G}_{BdG}$ matrix, we can rewrite the above equation as,
\begin{eqnarray}&&
    \begin{pmatrix} 0 &\mathcal{C}\\\nonumber\mathcal{C}&0
    \end{pmatrix} \mathcal{G}_{BdG}^T(k,t,t')
    \begin{pmatrix} 0 &\mathcal{C}^{-1}\\\mathcal{C}^{-1}&0
    \end{pmatrix}=\mathcal{G}_{BdG}(-k,t',t)\implies\\&&\nonumber
    \begin{pmatrix} 0 &\mathcal{C}\\\mathcal{C}&0
    \end{pmatrix} \begin{pmatrix}0& {G}_{BdG}(k,t,t')^T\\\tilde{G}_{BdG}^T(k,t,t')&0\end{pmatrix}
    \begin{pmatrix} 0 &\mathcal{C}^{-1}\\\mathcal{C}^{-1}&0
    \end{pmatrix}\\\nonumber&&\quad\quad= - \begin{pmatrix}0& \tilde{G}_{BdG}(-k,t',t)\\G_{BdG}(-k,t',t)&0\end{pmatrix}
    \end{eqnarray}
    We have thus reproduced the following relationship, even in the nonhermitian system,
    \beq
    \mathcal{C}G_{BdG}^T(k,t,t')\mathcal{C}^{-1}=-G_{BdG}(-k,t',t)\\
    \mathcal{C}\tilde{G}_{BdG}^T(k,t,t')\mathcal{C}^{-1}= -\tilde{G}_{BdG}(-k,t',t)\eeq
Diagonal blocks give us the definion of time-ordered product,
\be\langle T_tc_{k,a}(t)\tilde{c}^\dagger_{k,b}(t')\rangle=-\langle T_t\tilde{c}^\dagger_{k,b}(t')c_{k,a}(t)\rangle\ee
Again looking at the off-diagonal components, we have recovered the $SP^*OT^*$ rule,
\begin{equation}
    {\sf F}(k,t,t')=-{\sf F}^T(-k,t',t)
\end{equation}
We stress the fact that in the hermitian system, $SP^*OT^*$ rule is a direct consequence of the BdG-PHS while in the nonhermitian system, BdG-PHS is broken from the outset. In order to derive a $SP^*OT^*$ constraint, we have to introduce the biorthogonal basis. Hence, the full symmetry for the nonhermitian system is given by combinations of two symmetries: one stemming from the particle and hole excitations where the states are classified with respect to their transformation under $C$ and the second one, that still holds in the nonhermitian systems, name the $SP^*OT^*$ constraint. 

We thus introduce the extended symmetry for the nonhermitian systems where all pairing states can be classified as eigenstates of $C^*$ and $SP^*OT^*$. We call this symmetry a  $C^*\otimes(SP^*OT^*=-1)$. An immediate implication of this symmetry is that in nonhermitian systems,  multiple superconducting channels can open up as a consequence of broken particle-hole symmetry in the BdG Hamiltonian, where $C^* = \pm 1$. Yet with all these channels, the $SP^*OT^*$ constraint will be obeyed. 

\section{Unconventional superconductivity in model nonhermitian systems}
\label{secV}
In the previous sections, we gave a detailed analysis of the Berezinskii classification for the nonhermitian system.  Now, we give specific examples of such superconducting channels in the FSF setup shown in Fig. \eqref{fig:SFS}. As discussed in the sec-2, the spin-polarized current in the bulk of the superconductor gives rise to the finite life-time to the electrons in the superconductor. For demonstration purposes, we have assumed the bulk to be a BCS superconductor.
\begin{equation*}
    H_{BCS}(k)=\epsilon(k)\sum_{\sigma=\uparrow,\downarrow}c^\dagger_{k,\sigma}c_{k,\sigma}+\sum_{\sigma\ne\sigma'}\Delta(k)c^\dagger_{k,\sigma}c^\dagger_{-k,\sigma '}+h.c.
\end{equation*}
\begin{figure*}
    \includegraphics[scale=.35]{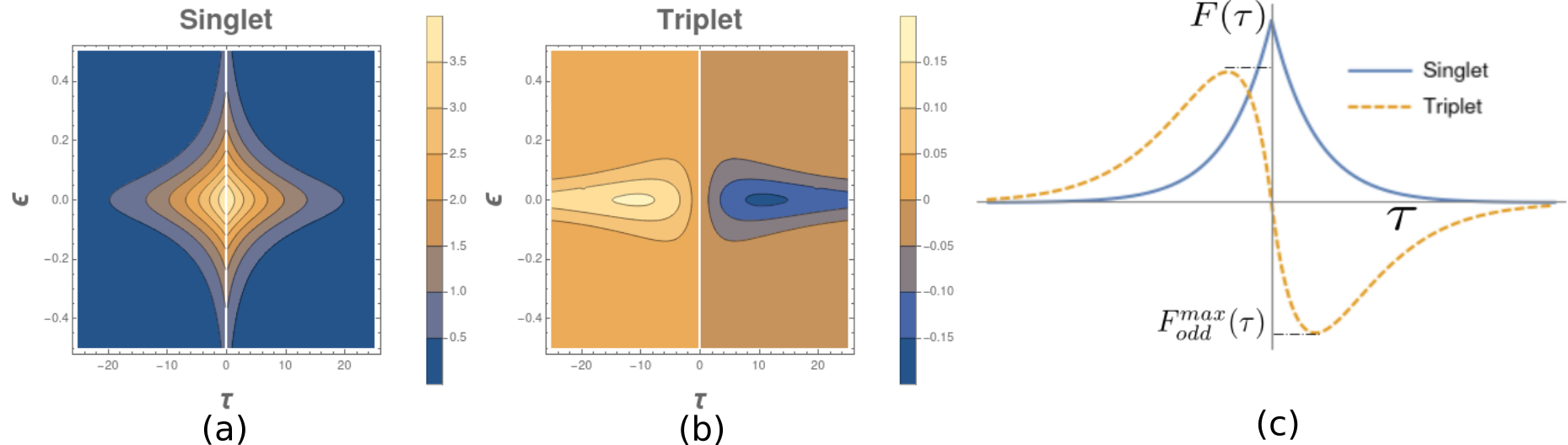}
    \caption{ $\Gamma_{\uparrow}=-\Gamma_{\downarrow}=\frac{\Delta}{10}$ (slow decay): (a-b) Anomalous pairing channels $F(\tau, \epsilon)$ with respect to imaginary time $\tau$ and onsite energy $\epsilon$. (c) $\epsilon=0$ cross sections of $F(\epsilon,\tau)$ from figure (a-b). We get two anomalous pairing channels: (a) ${\sf F}^{C^*}_{SP^*OT^*}={\sf F}^+_{-+++}$ (BCS spin singlet even frequency) (b) ${\sf F}^{C^*}_{SP^*OT^*}={\sf F}^+_{+++-}$ (Berezinskii spin triplet odd frequency). In the BCS channel, the anomalous paring amplitude takes maximum value when $\epsilon=\tau=0$. Berezinskii channel attains maximum at $\epsilon=0$, $\tau\sim \sqrt{\epsilon^2+\Delta^2}$.
    }
    \label{fig:comzman}
\end{figure*}

Where, $\epsilon(k)=v_F(k-k_F)$ and $\Delta(k)$ is assumed to be constant $\Delta$. $v_F$ and $k_F$ are Fermi velocity and Fermi momentum respectively. Now, once we turn on the finite lifetime of the electrons, in the bi-orthogonal basis the Hamiltonian will look like, 
\beq\nonumber
    &&H_{BCS}(k)=\epsilon(k)\sum_{\sigma=\uparrow,\downarrow}\Tilde{c}^\dagger_{k,\sigma}c_{k,\sigma}+\sum_{\sigma\ne\sigma'}\Delta(k)\tilde{c}^\dagger_{k,\sigma}\tilde{c}^\dagger_{-k,\sigma '}\\&&-\sum_{\sigma=\uparrow,\downarrow}i \Gamma_{\sigma}\Tilde{c}^\dagger_{k,\sigma}c_{k,\sigma}-\sum_{\sigma\ne\sigma'}\Delta^*(k){c}_{k,\sigma}{c}_{-k,\sigma '}
\eeq
Where $\Gamma_\sigma$ is the constant decay rate for the spin $\sigma=\uparrow,\downarrow$.
In this simple mean-field model, we can calculate the Green's function matrix. 
There will be a complex bandgap \cite{janik2001green, gongNonherm} in the system with band energies are given by, $E$, $-E^*$.  
\begin{equation}
    E=\sqrt{\Delta^2+(\epsilon(k)-i\Gamma)^2}+ i\gamma
\end{equation}
where $2\gamma=\Gamma_\downarrow-\Gamma_\uparrow$ and $2\Gamma=\Gamma_\uparrow+\Gamma_\downarrow$. $E$ and $E^*$ are complex conjugates of each other. Now using eq. (32), we can calculate the normal and anomalous Green's function to be,
\begin{equation}
    {\sf G}(\omega)_{\sigma\sigma}=\frac{\epsilon(k)-i\Gamma+\omega+i\sigma\gamma}{(\omega-i\sigma\gamma)^2-\Delta^2-(\epsilon(k)-i\Gamma)^2}
\end{equation}
\begin{equation}
    {\sf F}(\omega)_{\sigma\sigma'}=\frac{\sigma\Delta}{(\omega-i\sigma\gamma)^2-\Delta^2-(\epsilon(k)-i\Gamma)^2}
\end{equation}
To understand physical observables related to this nonhermitian BCS superconductor, we will start by constructing non-zero super-conducting anomalous pairing amplitudes in different cases of $\Gamma_\uparrow$ and $\Gamma_\downarrow$.

\subsection{Effective complex Zeeman field: $\Gamma_\uparrow=-\Gamma_\downarrow$ }
We will start our discussion with $\Gamma_\uparrow=-\Gamma_\downarrow$, i,e, spin-down particles are decaying out of the system, while spin-up particles are pumped into it. The total number of particles remains constant. This scenario essentially means $\Gamma=0$. 
\begin{equation}
    F(\omega)_{\uparrow\downarrow}=\frac{\Delta}{(\omega-i\gamma)^2-\Delta^2-\epsilon(k)^2}
\end{equation}
\begin{equation}
    F(\omega)_{\downarrow\uparrow}=-\frac{\Delta}{(\omega+i\gamma)^2-\Delta^2-\epsilon(k)^2}
\end{equation}

The effective imaginary complex field, $i\gamma$,  however, creates an essential disparity between $F(\omega)_{\uparrow\downarrow}$ and $F(\omega)_{\downarrow\uparrow}$. We can now define singlet, as well as triplet components,
\begin{equation*}
        {\sf F}_s={\sf F}_{\uparrow\downarrow}+{\sf F}_{\downarrow\uparrow},~~~F_t=F_{\uparrow\downarrow}-F_{\downarrow\uparrow}
\end{equation*}

\begin{figure}
    \centering
    \includegraphics[scale=.5]{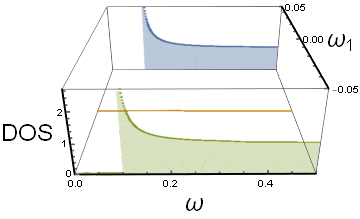}
    \caption{Density of states (DOS) $\rho_E$ with respect to real part ($\omega$) and imaginary part ($\omega_1$) of the frequency for the BCS type superconductor with spin decay, $\Gamma_\uparrow=-\Gamma_\downarrow=.5 \Delta$ (moderate decay). $\Delta$ is the superconducting gap in the BCS system. Spin up and down starts with a similar profile as BCS superconductor at different values of $\omega_1$. Spin up DOS is picked up at $\omega_1=\Gamma_\uparrow$ and decaying out of the system. Spin down DOS is picked up at $\omega_1=\Gamma_\downarrow$ and pumped into the system. This spin imbalance gives rise to the triplet SC component. This spin imbalance occurs from the dynamics of the system and different from the spin imbalance in a static system under a Zeeman field.}
    \label{fig:dos01}
\end{figure}

These pairing states respectively have $F_s$: $S = -1, P^* = O = T^* =1$ ; $F_t$:$S = O = P^* = 1, T^* = -1$ and  $SP^*OT^*=-1$. Singlet and triplet states have opposite $T^*$ symmetry. As shown in Fig. \eqref{fig:comzman}, one indeed  gets an odd-frequency Berezinskii pairing amplitude in this case. The pairing amplitudes, however, are even under particle-hole parity ($\epsilon\leftrightarrow-\epsilon$) (Fig. \eqref{fig:comzman}). We can classify these two order parameters according to Table \ref{tab:class},
 \be{\sf F}_s={\sf F}^+_{-+++},\quad{\sf F}_s={\sf F}^+_{+++-}\ee
 The form of the anomalous Green's function is similar to the BCS superconductor under the Zeeman field, $h=i\Gamma$\cite{linder2015superconducting}. This current system, however, describes a dynamic system rather than a static one. The density of states profile (see Fig.\eqref{fig:dos01}) of our system explicitly points out to such a distinction.
\begin{figure*}
    \centering
    \includegraphics[scale=.4]{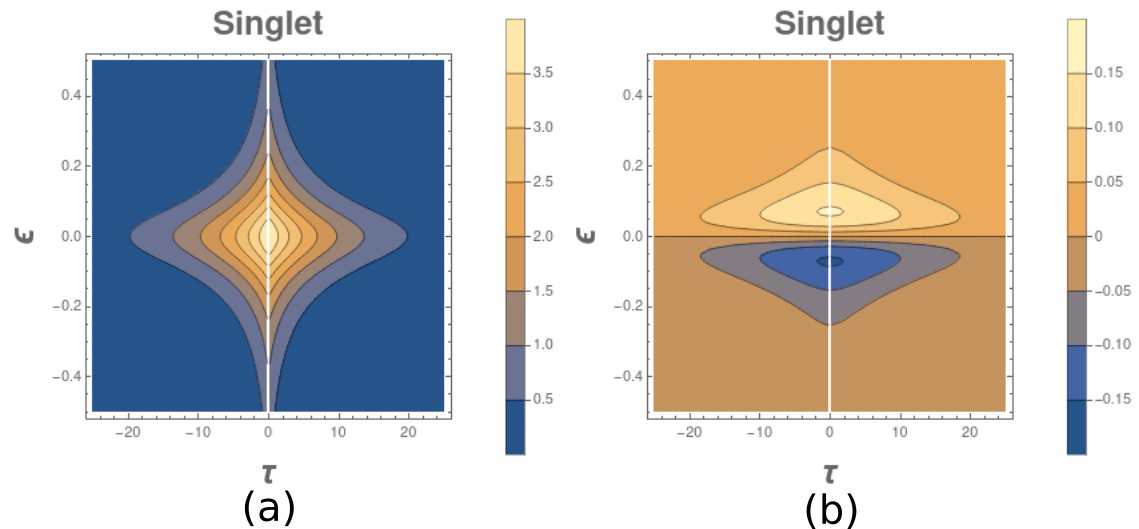}
    \caption{ $\Gamma_{\uparrow}=\Gamma_{\downarrow}=\frac{\Delta}{10}$ (slow decay): (a-b) Anomalous pairing channels $F(\tau, \epsilon)$ with respect to imaginary time $\tau$ and onsite energy $\epsilon$. We get two anomalous pairing channels: (a) ${\sf F}^{C^*}_{SP^*OT^*}={\sf F}^+_{-+++}$ (BCS, spin-singlet, even-frequency) (b) ${\sf F}^{C^*}_{SP^*OT^*}={\sf F}^-_{-+++}$ (Odd $\epsilon$, spin-singlet, even-frequency). Both of these channels satisfy $SP^*OT^*=-1$ identically but gets different sign under the action of $C^*$.}
    \label{fig:comchem}
\end{figure*}
\begin{figure}
    \centering
    \includegraphics[scale=.5]{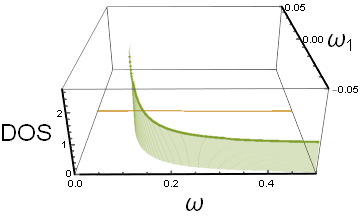}
    \caption{Density of states (DOS) $\rho_E$ with respect to real part ($\omega$) and imaginary part ($\omega_1$) of the frequency for the BCS type superconductor with spin decay, $\Gamma_\uparrow=\Gamma_\downarrow=.5 \Delta$ (moderate decay). Up and down spins are decaying unanimously. This results in the shrinking of the Fermi-surface and renormalizing the DOS over time.}
    \label{fig:dos02}
\end{figure}
\subsection{Complex chemical potential case: $\Gamma_{\uparrow}=\Gamma_{\downarrow}$}
In the case $\Gamma_{\uparrow}=\Gamma_{\downarrow}$, we get particles are decaying out of the system at a constant rate, $\Gamma$. This situation can be thought of constant shrinking of Fermi surface or imaginary chemical potential. In this scenario, we have $\gamma=0$,
\begin{equation}
    {\sf F}(\omega)_{\uparrow\downarrow}=-{\sf F}(\omega)_{\downarrow\uparrow}=\frac{\Delta}{\omega^2-\Delta^2-(\epsilon(k)-i\Gamma)^2}
\end{equation}
As there is no spin dependence in the potential, we expect the spin-up, spin-down channel pairing amplitude to be equal to the spin-up channel. This results in no spin-triplet channel in the system. The complex chemical potential, however, gives rise to the constant leaking of the system and creates a particle-hole imbalance in the system. To study this particle-hole imbalance, we should look into ${\sf F}(\omega)_{\sigma\sigma'}$ as well as its conjugate $\tilde{\sf F}(\omega)_{\sigma\sigma'}$,
\begin{equation}
    \tilde{\sf F}(\omega)_{\sigma\sigma'}=C^*{\sf F}(\omega)_{\sigma\sigma'}=\frac{\Delta_\sigma}{\omega^2-\Delta^2-(\epsilon(k)+i\Gamma)^2}
\end{equation}
Where, $\sigma\ne\sigma'$ and  $\Delta_\uparrow=-\Delta_\downarrow=\delta$. In order to separate the odd $\epsilon$ channel from the even $\epsilon$ channel, we define,
\be {\sf F}^e={\sf F}+\tilde{\sf F}; \quad {\sf F}^o={\sf F}+\tilde{\sf F}; \ee
As shown in Fig. \eqref{fig:comchem}(a), ${\sf F}^e$ is the normal single orbital, spin singlet BCS channel and denoted by ${\sf F}^+_{-+++}$ according to Table \ref{tab:class}. ${\sf F}^o$ in Fig. \eqref{fig:comchem} is odd $\epsilon$ single orbital, spin singlet, even frequency state. Similar state has been studied previously by Mila and Abraham\cite{PhysRevLett.67.2379} in hermitian set up. Our state has been realized in a dynamic system with decaying density of states as shown in Fig. \eqref{fig:dos02}.   

\begin{figure}
    \centering
    \includegraphics[scale=.5]{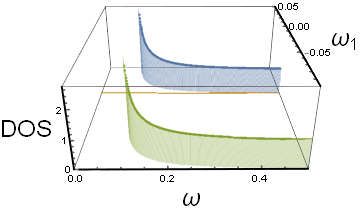}
    \caption{Density of states (DOS) $\rho_E$ with respect to real part ($\omega$) and imaginary part ($\omega_1$) of the frequency for the BCS type superconductor with spin-polarized decay, $\Gamma_\uparrow=.5 \Delta$ (moderate decay), $\Gamma_\downarrow=0$. Up spins are decaying out of the system and shrinking the overall Fermi-surface. Spin-up DOS is shifting away from the $\omega$ axis for a higher frequency, while spin-down DOS is shifting towards $\omega$ axis.}
    \label{fig:dos11}
\end{figure}
\begin{figure*}
    \centering
    \includegraphics[scale=.4]{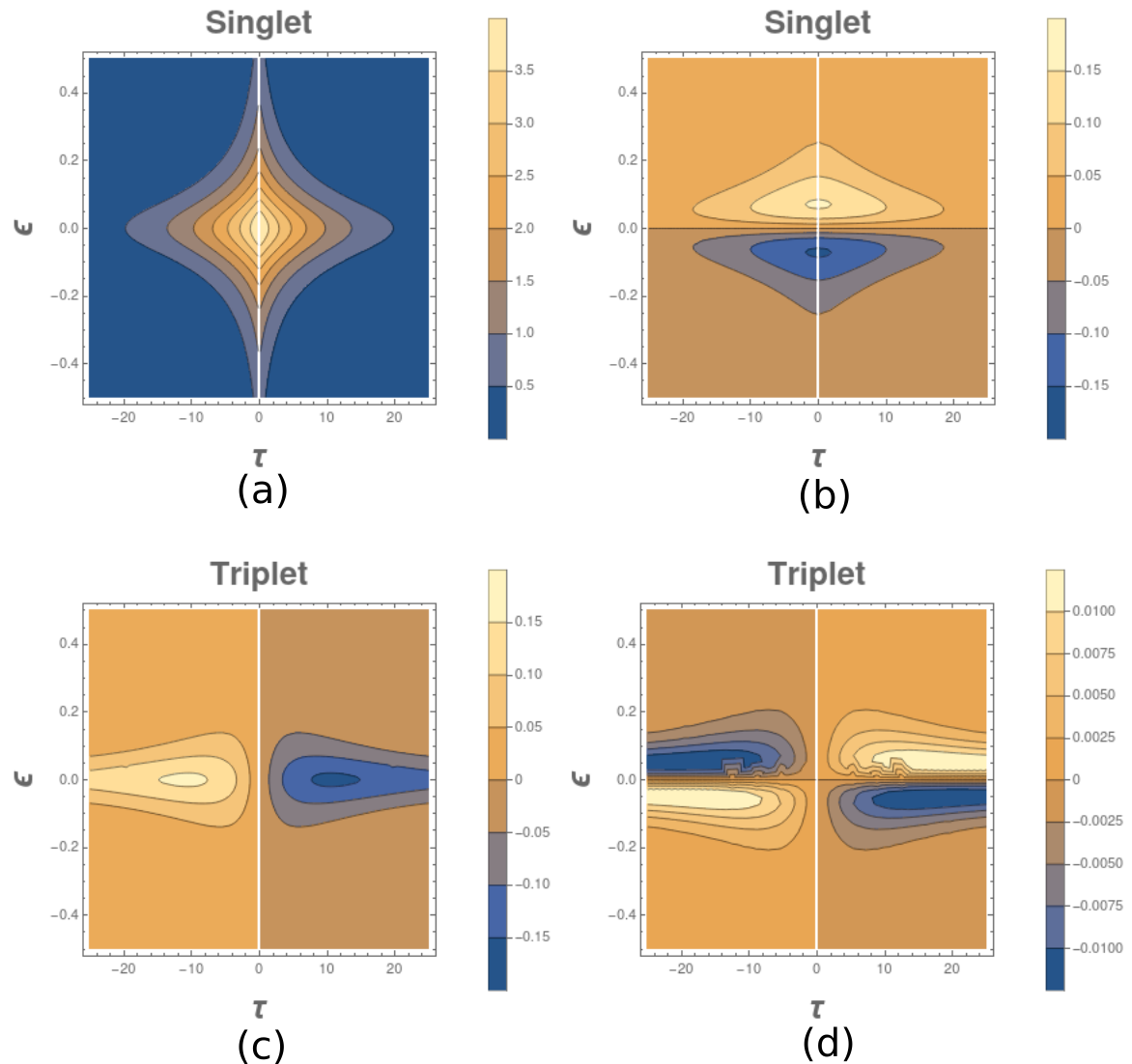}
    \caption{ $\Gamma_{\downarrow}=0,\Gamma_{\downarrow}=\frac{\Delta}{10}$ (slow decay) : (a-d) Anomalous pairing channels $F(\tau, \epsilon)$ with respect to imaginary time $\tau$ and onsite energy $\epsilon$. We get four anomalous pairing channels: (a) ${\sf F}^{C^*}_{SP^*OT^*}={\sf F}^+_{-+++}$ (BCS spin singlet even frequency) (b) ${\sf F}^{C^*}_{SP^*OT^*}={\sf F}^+_{+++-}$ (Berezinskii spin triplet odd frequency) (c) ${\sf F}^{C^*}_{SP^*OT^*}={\sf F}^-_{-+++}$ (Odd $\epsilon$ spin singlet even frequency) (d) ${\sf F}^{C^*}_{SP^*OT^*}={\sf F}^-_{+++-}$ (Odd $\epsilon$ spin triplet odd frequency). (a-b) are even while (c-d) are odd under particle-hole exchange.}
    \label{fig:nhsc}
\end{figure*}

\subsection{Spin polarized decay ($\Gamma_{\downarrow}=0$)}

All of the above cases can be seen once we study the decay of only one kind of spin. This scenario can be realized in the FSN structure, where a bulk superconductor is coupled to a ferromagnet and a normal metal in the opposite end and a small voltage difference is applied from ferromagnet to the normal metal. A small but unpolarized current will enter the superconductor from the metal, while a spin-polarized current will pass through the superconductor-ferromagnet interface. Due to spin decay, the Fermi surface is continuously shrinking and creates a particle-hole imbalance. Moreover, the density of states profile for spin-up and down states are separated in the frequency plane and create spin-imbalance in the dynamic system (see Fig. \eqref{fig:dos11}). In this case, we have $\gamma=-\Gamma$ and can construct four different anomalous paring channels as shown in Fig. \eqref{fig:nhsc},

\begin{eqnarray}
    {\sf F}^e_t={\sf F}_{\sigma\sigma'}+\tilde{\sf F}_{\sigma'\sigma}+{\sf F}_{\sigma'\sigma}+\tilde{\sf F}_{\sigma\sigma'}\\
    {\sf F}^e_s={\sf F}_{\sigma\sigma'}+\tilde{\sf F}_{\sigma'\sigma}-{\sf F}_{\sigma'\sigma}-\tilde{\sf F}_{\sigma\sigma'}\\
    {\sf F}^o_t={\sf F}_{\sigma\sigma'}-\tilde{\sf F}_{\sigma'\sigma}+{\sf F}_{\sigma'\sigma}-\tilde{\sf F}_{\sigma\sigma'}\\
    {\sf F}^o_s={\sf F}_{\sigma\sigma'}-\tilde{\sf F}_{\sigma'\sigma}-{\sf F}_{\sigma'\sigma}+\tilde{\sf F}_{\sigma\sigma'}
\end{eqnarray}
According to our $C^*$ and $SP^*OT^*$ classification in the Table \ref{tab:class}, we can explicitly write down the quantum numbers of the above pairing amplitude channels.
\begin{eqnarray}
    {\sf F}^e_t:{\sf F}^+_{+++-};\quad {\sf F}^e_s:{\sf F}^+_{-+++};\\
    {\sf F}^o_t:{\sf F}^-_{+++-};\quad {\sf F}^0_s:{\sf F}^-_{-+++};
\end{eqnarray}
We have previously discussed three of the above channels, shown in Fig. \eqref{fig:nhsc}(a-c). In the spin-polarized decay we have obtained a new odd frequency channel, ${\sf F}^o_t$ (Fig. \eqref{fig:nhsc}(d)). This channel is also odd under particle-hole exchange. Such superconductivity has not been discussed in the literature to our best knowledge. 

The relative amplitudes of these SC channels can be modified by changing the decay rate. In our examples, we have chosen decay rates to be smaller than $\Delta$. One can, however, choose a very different set of decay rates to achieve different profiles of ${\sf F}(\epsilon,\tau)$, but their $C^*$, $S$, $P^*$, $O$ and $T^*$ quantum numbers remain unchanged.  
\section{Spin susceptibility in nonhermitian superconductors }
\label{secVI}
We have, so far, given classification schemes nonhermitian superconductors and realized a few of them in the BCS type framework. These superconductors not only have a very distinct signature in the DOS profile, but we also expect very distinct thermodynamic properties of them. To substantiate this point, here we have calculated the spin susceptibility of nonhermitian systems. Spin susceptibility $\chi_{zz}(k,\omega)$ at temperature $T$ is defined in terms of spin operators \cite{mahan2013many},
\begin{figure}
    \centering
    \includegraphics[scale=0.5]{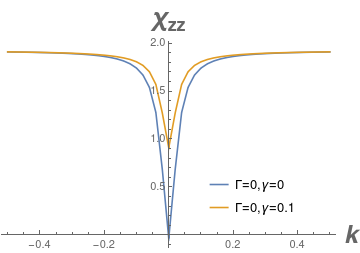}
    \caption{Static spin susceptibility with respect to momentum ($k$) for the nonhermitian BCS superconductor ($\Delta=.01$) with Zeeman type spin decay (fast decay, $\gamma=10\Delta$) is compared to the isolated/ normal BCS superconductor case. Both of them for large $k$ (far from Fermi surface) shows paramagnetic behavior. Near $k=0$, BCS spin-susceptibility falls sharply to zero and shows diamagnetism. The spin-triplet part of the nonhermitian BCS superconductor has non-zero spin susceptibility at $k=0$, as shown in the figure.}
    \label{fig:my_label}
\end{figure}
\begin{figure}
    \centering
    \includegraphics[scale=0.5]{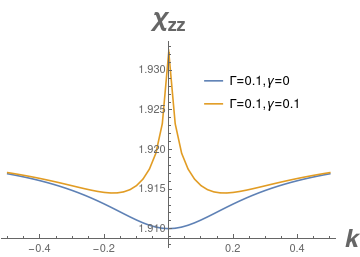}
    \caption{Static spin susceptibility with respect to momentum ($k$) is compared for the nonhermitian BCS superconductor ($\Delta=.01$) with complex chemical potential and another nonhermitian BCS superconductor with spin-polarized decay (fast decay, $\Gamma=\gamma=10\Delta$). Both of them behaves like a paramagnetic material with a small decrease near $k=0$. This observation suggests a faint diamagnetic behavior compared to an isolated BCS superconductor. For spin-polarized decay, we have an additional pick at $k=0$ (near Fermi surface). We associate this pick with the triplet superconducting channels in the system.}
    \label{fig:my_label1}
\end{figure}

\be\nonumber\chi_{zz}(k,\omega)=4\mu_B^2\int\limits_0^{\frac{1}{K_BT}}\sum_{k, \omega_n}d\tau e^{i\omega_n\tau}\langle T_\tau S_z(k,\tau)S_z(-k,0)\rangle\ee
Where $\tau$ and $\omega_n$ are Matsubara time and Matsubara frequency respectively. $\mu_B$ is Bohr magneton, and $K_B$ is Boltzman constant. In hermitian systems, $S_z(q,\tau)$ , in terms of Pauli matrix $\sigma^z$, is given by, 
\be S_z(k,\tau)=\sum_{k,\sigma=\uparrow,\downarrow} \sigma^z_{\sigma\sigma'}c^\dagger_{q+k,\sigma}(\tau)c_{q,\sigma'}(0)\ee
We can can define a similar quantity in the nonhermitian system, 
\beq\nonumber S_z(k,\tau)=\frac{1}{2}\sum_{q,\sigma=\uparrow,\downarrow} \sigma^z_{\sigma\sigma'}(\tilde{c}^\dagger_{q+k,\sigma}(\tau)c_{q,\sigma'}(0)\\+{c}^\dagger_{q+k,\sigma}(\tau)\tilde{c}_{q,\sigma'}(0))\eeq
Where, $q$ and $k$ are momentum vector and $q+k$ stands for the vector sum. Hence, in a nonhermitian system, we can write spin-susceptibility as a sum of four 4-point functions,
\beq\nonumber
\chi_{zz}(k,\omega)=\mu_B^2\sum_{q,q',\sigma,\sigma'}\sigma^z_{\sigma\sigma}\sigma^z_{\sigma'\sigma'}\int\limits_0^{\frac{1}{K_BT}}d\tau e^{i\omega\tau}\\\nonumber
(\langle T_\tau \tilde{c}^\dagger_{q+k,\sigma}(\tau){c}_{q,\sigma}(\tau)\tilde{c}^\dagger_{q'-k,\sigma'}(0){c}_{q',\sigma'}(0)\rangle\\\nonumber+\langle T_\tau {c}^\dagger_{q+k,\sigma}(\tau)\tilde{c}_{q,\sigma}(\tau){c}^\dagger_{q'-k,\sigma'}(0)\tilde{c}_{q',\sigma'}(0)\rangle\\\nonumber+\langle T_\tau \tilde{c}^\dagger_{q+k,\sigma}(\tau){c}_{q,\sigma}(\tau){c}^\dagger_{q'-k,\sigma'}(0)\tilde{c}_{q',\sigma'}(0)\rangle\\+\langle T_\tau {c}^\dagger_{q+k,\sigma}(\tau)\tilde{c}_{q,\sigma}(\tau)\tilde{c}^\dagger_{q'-k,\sigma'}(0){c}_{q',\sigma'}(0)\rangle)
\eeq
In the simple nonhermitian models, we were discussing so far, the only nontrivial contributions come from, 
\beq\nonumber&& 
{\sf{G}}(k,\tau)_{\sigma\sigma'}=\delta_{\sigma\sigma'}\langle T_\tau \tilde{c}^\dagger_{k,\sigma}(\tau){c}_{k,\sigma'}(0)\rangle,
\\\nonumber&&
\tilde{\sf{G}}(k,\tau)_{\sigma\sigma'}=\delta_{\sigma\sigma'}\langle T_\tau {c}^\dagger_{k,\sigma}(\tau)\tilde{c}_{k,\sigma'}(0)\rangle, 
\\\nonumber&&
{\sf{F}}(k,\tau)_{\sigma\sigma'}=(1-\delta_{\sigma\sigma'})\langle T_\tau {c}_{-k,\sigma}(\tau){c}_{k,\sigma'}(0)\rangle,
\\\nonumber&&
\tilde{\sf{F}}(k,\tau)_{\sigma\sigma'}=(1-\delta_{\sigma\sigma'})\langle T_\tau \tilde{c}_{-k,\sigma}(\tau)\tilde{c}_{k,\sigma'}(0)\rangle, 
\\\nonumber&&
{\sf{F}}^\dagger(k,\tau)_{\sigma\sigma'}=(1-\delta_{\sigma\sigma'})\langle T_\tau {c}^\dagger_{k,\sigma}(\tau){c}^\dagger_{-k,\sigma'}(0)\rangle,
\\\nonumber&&
\tilde{\sf{F}}^\dagger(k,\tau)_{\sigma\sigma'}=(1-\delta_{\sigma\sigma'})\langle T_\tau \tilde{c}^\dagger_{k,\sigma}(\tau)\tilde{c}^\dagger_{-k,\sigma'}(0)\rangle. 
\eeq
With the above assumption, Pauli spin-susceptibility does have a simpler form:
\beq\nonumber
\chi_{zz}(k,\omega)=\frac{\mu_B^2}{K_B T}\Re \{\sum_{q,\tilde{\omega}_n,\sigma\sigma'}\\\nonumber{\sf G}_{\sigma\sigma'}(q, i\tilde{\omega}_n){\sf G}_{\sigma'\sigma}(q+k, i\tilde{\omega}_n+\omega)+\\{\sf F}_{\sigma\sigma'}(q, i\tilde{\omega}_n)\tilde{\sf F}^\dagger_{\sigma'\sigma}(q+k, i\tilde{\omega}_n+\omega)\}
\eeq
In the case of fast decay ($\delta<<$ decay rate), Pauli spin susceptibility indeed showed non-trivial features as discussed in Fig. \eqref{fig:my_label}-\eqref{fig:my_label1}.

\section{Conclusions}\label{conclusion}
This work represents a detailed look into possible symmetries and constraints one would have in developing pairing states in nonhermitian systems.
We extended the classification of the pairing states to nonhermitian systems. Nonhermitian terms in the Hamiltonian provide new dynamic contributions and naturally lead to odd frequency pairing correlations that develop in the bulk of the material as a result of damping. We find that paring states obey the $SP^*OT^*$ classification and all states possess definite spatial and time parities as in equilibrium. The new aspect of nonhermitian dynamics is the emergence of the processes that convert electrons into holes. Nonhermitian terms differentiate between the hole creation and electron annihilation processes. As a result, we define the new symmetry  $C = \pm 1$ that operates on the space of nonhermitian systems in the presence of the damping.   This symmetry allows for new pairing states that are the so-called odd- energy states, considered a long time ago \cite{cohen1964coherent,nakajima1964superconductivity,PhysRevLett.67.2379}. We now find that these odd-energy states are the new class of solutions with $C = -1$ and should be expected to coexist with BCS state in nonhermitian superconducting systems. 

We illustrate our general approach to a new states generated in a bulk BCS superconducting state subject to the spin dependent damping. Within this simple approach we find new possible  pairing states developing including OFSc and odd energy states. One can take this approach to any other,  physically motivated non-BCS pairing states and likely find similar plethora of new pairing states that emerge as a consequence of nonhemitian dynamics.  We thus are confident that nonhermitian dynamics provides an exciting new route to probe exotic superconducting states including Berezinskii pairing. Since our classification permits extenal pumping we expect many of these states emerge in dynamically driven superconducting materials. 
\begin{acknowledgments}
We are grateful to D. Kuzmanovski and J. Cayao for useful discussions. 
This work was supported by the VILLUM FONDEN via the Centre of Excellence for Dirac Materials (Grant No. 11744) and the European Research Council under the European Unions Seventh Framework ERS-2018-SYG 810451 HERO.
\end{acknowledgments}
%
\end{document}